# Perception and Utilization of Data Carpentry Tools among Research Scholars at University of Kerala

**Jagan B L**
*Research Scholar*
*Dept. of Library and Information Science,*
*University of Kerala, Thiruvananthapuram*
*https://orcid.org/0009-0007-2295-797X*
*jaganbl2024@keralauniversity.ac.in*

**Dr. Sudhi S. Vijayan**
*Assistant Professor & Head*
*Dept. of Library and Information Science,*
*University of Kerala, Thiruvananthapuram*
*https://orcid.org/0000-0002-2228-9031*
*sudhivijayan@keralauniversity.ac.in*

**Dr. V. R. Renjith**
*Assistant Librarian*
*Kerala University Library, University of Kerala,*
*Thiruvananthapuram, Kerala*
*https://orcid.org/0000-0002-2389-7338*
*renjithlib@keralauniversity.ac.in*

## Abstract

Data Carpentry (DC) has been developed to instruct participants on essential concepts, skills, and tools for more effective and reproducible data management. DC emphasizes data literacy, aiming to equip researchers with the skills necessary to retrieve, view, manipulate, analyze, and store their own and other's data in an open and reproducible manner to derive insights from data. The objective of this study was to assess the perceptions and attitudes towards DC among research scholars of University of Kerala and to find out the purpose for which DC tools is used by the research scholars in the University of Kerala. This study requires a sample of 167 and 179 research scholars completed a questionnaire from a total population of 294 using the Krejcie and Morgan formula, with a 95% confidence interval and a 5% margin of error. SPSS version 22 was used to perform descriptive and inferential statistics on the data. 65.9% of research researchers have experience with DC tools. 48.6% of research researchers at the University of Kerala use DC tools primarily for data analysis. 40.2% of research academics identified insufficient access to training as the most significant obstacle they experience while utilizing DC tools in the study. DC emphasizes data literacy, aiming to equip researchers with the skills necessary to access, view, modify, analyze, and manage their own and others' data in an open and repeatable manner to derive insights from data.

## 1. Introduction

The exponential growth of digital data in various disciplines has necessitated the development of specialized tools and techniques for data management analysis and visualization. Data Carpentry (DC), an initiative that emerged from the software carpentry movement aims to provide researchers with the skills and knowledge required to effectively work with data. Thorough research with high quality data influences the development of future generations, society, and the nation at large. (Das, 2017) Proper training enables researchers to retrieve, view, manipulate, analyze and store their and other's data in an open and reproducible way in order to extract knowledge from data (Teal, et al. 2015). Data Carpentry's focus is on the introductory computational skills needed for data management and analysis in all domains of research. (Phlow & Jekyll, 2025). DC skill sets are essentially domain-specific and naturally varies widely from discipline to discipline. (Mukhopadhyay, 2021).

The University of Kerala, as a premier institute of higher learning has a large community of research scholars who generate and utilize significant amount of data in their research endeavors. This study investigates the perception and utilization of DC tools among Research Scholars (R/S) at the University of Kerala, with a focus on understanding the current state of data management practices, the awareness and adoption of DC tools and the challenges faced by the researchers in effectively utilizing these tools. The findings of the study will provide valuable insights into the data management needs of research scholars and inform the development target training programmes and support services to enhance their data management skills.

## 2. Objectives

1) To examine the awareness of DC among the R/S in the University of Kerala.
2) To find out the perception of DC tools on various aspects among the R/S in the University of Kerala.
3) To find out the purpose for which DC tools is used by the R/S in the University of Kerala.
4) To know the challenges faced by R/S while using DC tools in connection with their research.

## 3. Hypotheses

1) Opinions regarding the perception of DC tools of all statements are equal to average level.
2) There is no significant difference between Gender with regard to all the statements of DC tools perception
3) Level of perception of DC tools are equally distributed among R/S
4) There is no association between R/S faculty of subjects and level of perception of DC tools.

## 4. Methodology

This study used a survey research approach to assess the perception and use of DC tools among R/S at the University of Kerala. The study focused on specific departments where DC is applied. These departments have been selected through purposive sampling, and the sample size has been determined using the Krejcie and Morgan Table (1970). From the total population of 294 researchers, a sample size of 167 is required. A self-designed questionnaire was sent to 200 research researchers using Google Forms as a survey tool. Of the 200 questionnaires distributed, 179 have been completed properly. The sample size proportion is determined by the number of scholars in each department, as presented in Table 1. The final sample included 179 R/S from the below mentioned departments.

**Table 1: Department and Sample Size**

| Department | Sample Received | Sample Required |
|---|---|---|
| Bioinformatics | 5 | 3 |
| Botany | 26 | 22 |
| Chemistry | 9 | 9 |
| Computer Science | 11 | 11 |
| Economics | 9 | 9 |
| Environmental Science | 4 | 4 |
| Geology | 13 | 13 |
| Library and Information Science | 9 | 5 |
| Management | 9 | 9 |
| Physics | 14 | 14 |
| Political Science | 31 | 31 |
| Psychology | 12 | 11 |
| Sociology | 13 | 13 |
| Zoology | 14 | 13 |
| **Total** | **179** | **167** |

## 5. Review of Literature

Mukhopadhyay, P., & Mukhopadhyay M. (2024) highlight open data in libraries and data management and analysis abilities for library workers in their research. Library carpentry focuses on data curation, textual data management, and bibliographic content negotiation. Library carpentry provides these methods to better data-driven information. The article highlights case examples on how library carpentry methods like bibliographic content negotiation, regular expressions, named entity recognition, machine translation, and data reconciliation may help libraries adapt to the changing information world and better serve their customers. (Mukhopadhyay, P., & Mukhopadhyay M, 2024).

Mukhopadhyay, P et al (2021) say library and information science is always changing and LIS practitioners are tech-savvy. A novel LIS solution that applies data science to bibliographies is evaluated in this research. Library carpentry manipulates bibliographic databases. After explaining library carpentry principles, three case studies show its components and approaches. Case studies demonstrate real-world problem-solving using Openrefine and open datasets. Case studies (I) apply library carpentry to e-book selection using socio-academic web space data, (II) show how to quickly get an overview of institutional contributions to open access, and (III) show gender analysis using a name-to-gender inference service and data wrangling. Each library carpentry case study solves real-world processional difficulties with a rich and realistic dataset. (Mukhopadhyay P, 2021)

According to Seidlmayer et al. (2020), libraries have evolved into sophisticated information hubs. This progress demands computer skills and a new learning mentality. “Library Carpentry” from “The Carpentries” community teaches computational principles and techniques thoroughly and empirically. ZB MED-Information Centre for Life Sciences promotes digital literacy using Library Carpentry. (Seidlmayer, E, 2020).

## 6. Data Analysis and Interpretation

The collect data was analyzed, tabulate, and interpret with the help of Microsoft Excel and Statistical software SPSS Version 22.

### 6.1 Background Details of Respondents

Table 2 gives the background details of the respondents.

**Table 2: Background Details of Respondents**

| Sl. No. | Attribute | Variables | Number (%) |
|---|---|---|---|
| 1 | Gender | Male | 60 (33.5) |
| | | Female | 119 (66.5) |
| 2 | Course of study | Ph.D | 173 (96.6) |
| | | PDF | 6 (3.4) |
| 3 | Generation | Y | 108 (60.3) |
| | | Z | 71 (30.7) |
| 4 | Faculty | Science | 96(53.6) |
| | | Social Science | 83(46.4) |
| 5. | Familiarity in DC tools | Familiar | 118(65.9) |
| | | Not Familiar | 61(34.1) |
| 6. | Recommendation of DC Tools to peers | Recommended | 172(96.1) |
| | | Not Recommended | 7(3.9) |
| 7. | Usage Frequency of Data Carpentry Tools | Daily | 9 (5.0) |
| | | Weekly | 24 (13.4) |
| | | Occasionally | 97 (54.2) |
| | | Rarely | 49 (27.4) |
| 8. | Purpose of Data Carpentry Tools* | Data Cleaning | 25 (13.9) |
| | | Data Analysis | 87 (48.6) |
| | | Data Visualization | 74 (41.3) |
| | | Reproducible Research | 36 (20.1) |

*Multiple Choice

Among the 179 respondents, 119 (66.5%) are female and 60 (33.5%) are male. This implies that female researchers are more when compared to male. It is evident from the result that majority of respondents are Ph.D. scholars, comprising 96.6%, while only 3.4% are post-doctoral fellows.

Individuals who were born within the same time frame as a generation. Millennials, or Generation Y (Gen Y) is defined as those born between 1980 and 1996. The respondents of Generation Z (Gen Z) were born between the years 1997 and 2012. Zoomers and post-millennials are other names for this generation. Of the 179 responders, 108 (60.3%) belong to Generation Y, and rest of the 71 (30.7%) are in Generation Z. A substantial percentage of researchers at Kerala University are aged between 28 and 44.

Faculty wise analysis reveals that more than half of the respondents are from Science faculty 96 (53.6) and the rest of the 83 (46.4) Social Science faculty. In response to the query regarding familiarity with data carpentry, 65.9% of researchers indicated that they are familiar, while the remaining 34.1% reported that they are not familiar. The result indicates that the lion share of the respondents (96.1%) would recommend DC tools to theirs peers and 3.9% of the respondents would not recommend DC tools to their peers.

A considerable portion of respondents (58.7%) used DC tools occasionally, whereas 27.4% researchers used DCT rarely. 13.4% of researchers used it once a week, next in line. Just 5% of researchers utilized it daily.

Nearly half of the respondents (48.6%) selected data analysis as the major application of DC tools, with a large fraction using these tools for data visualization (41.3%) successfully. However, 20.1% were utilized for reproducible research and 13.9% for data cleaning. **(Objective 3)**

## 6.2 Perception and Utilization

The rest of the analysis was done based on the perception and utilization of DC tools.

### 6.2.1 Prior Experience with DC Tools

Researchers often benefit from past expertise using data construction tools. These technologies, which simplify data organization, cleaning, and analysis, assist to streamline processes and increase the efficiency of dealing with enormous datasets. It need not just technical expertise, but also confidence, to address complex datasets in a logical and effective manner.

**Table 3: Prior Experience with Data Carpentry Tools**

| Experience | No. of Researchers | % |
|---|---|---|
| None | 36 | 20.1 |
| Beginner | 61 | 34.1 |
| Intermediate | 82 | 45.8 |
| Advanced | 0 | 0.0 |
| **Total** | **179** | **100.0** |

According to the data presented in Table 3, prior experience with the DC tools indicates that 45.8% of researchers are at an intermediate level, while 34.1% are categorized as beginners. Simultaneously, 20.1% lack experience with DC tools. It is noteworthy that none of the researchers are at an advanced level of DC tools.

### 6.2.2 Way of knowing about Data Carpentry Tools

The following are the some of the possible ways by which the R/S came to know about DC tools. This question is applicable only to those respondents who are familiar with DC tools.

**Table 4: Way of knowing about Data Carpentry Tools**

| Way of knowing | No. of Researchers | % |
|---|---|---|
| Colleagues/Peers | 47 | 39.8 |
| Online Resources | 61 | 51.7 |
| University Events | 10 | 8.5 |
| **Total** | **118** | **100.0** |

The data presented in Table 4 indicates that 51.7% of researchers acquired knowledge of DC tools through online resources, while 39.8% gained insights with assistance from colleagues or peers. A mere 8.5% of researchers acquire knowledge through university events.

### 6.2.3 Relevance of Data Carpentry Tools

DC tools are essential in the contemporary data-centric environment, especially within the research domain. These DC tools enhance the effective interpretation of extensive datasets, fostering improved decision-making and knowledge generation.

**Table 5: Relevance of DCT**

| Opinion | No. of Researchers | % |
|---|---|---|
| Not Relevant | 3 | 1.7 |
| Somewhat Relevant | 22 | 12.3 |
| Neutral | 24 | 13.4 |
| Relevant | 99 | 55.3 |
| Very relevant | 31 | 17.3 |
| **Total** | **179** | **100.0** |

The data presented in Table 5 indicates that 55.3% of researchers concur that DC tools is relevant, while 17.3% believe it is very relevant. Moreover, 13.4% responded neutrally.

### 6.2.4 Perception of DC Tools

DC tools are frequently regarded as vital assets for optimizing data-related activities, especially among researchers, analysts, and individuals managing sophisticated datasets. These tools streamline the complex processes of data cleaning, transformation, and visualization, enhancing accessibility and efficiency for users across various skill levels.

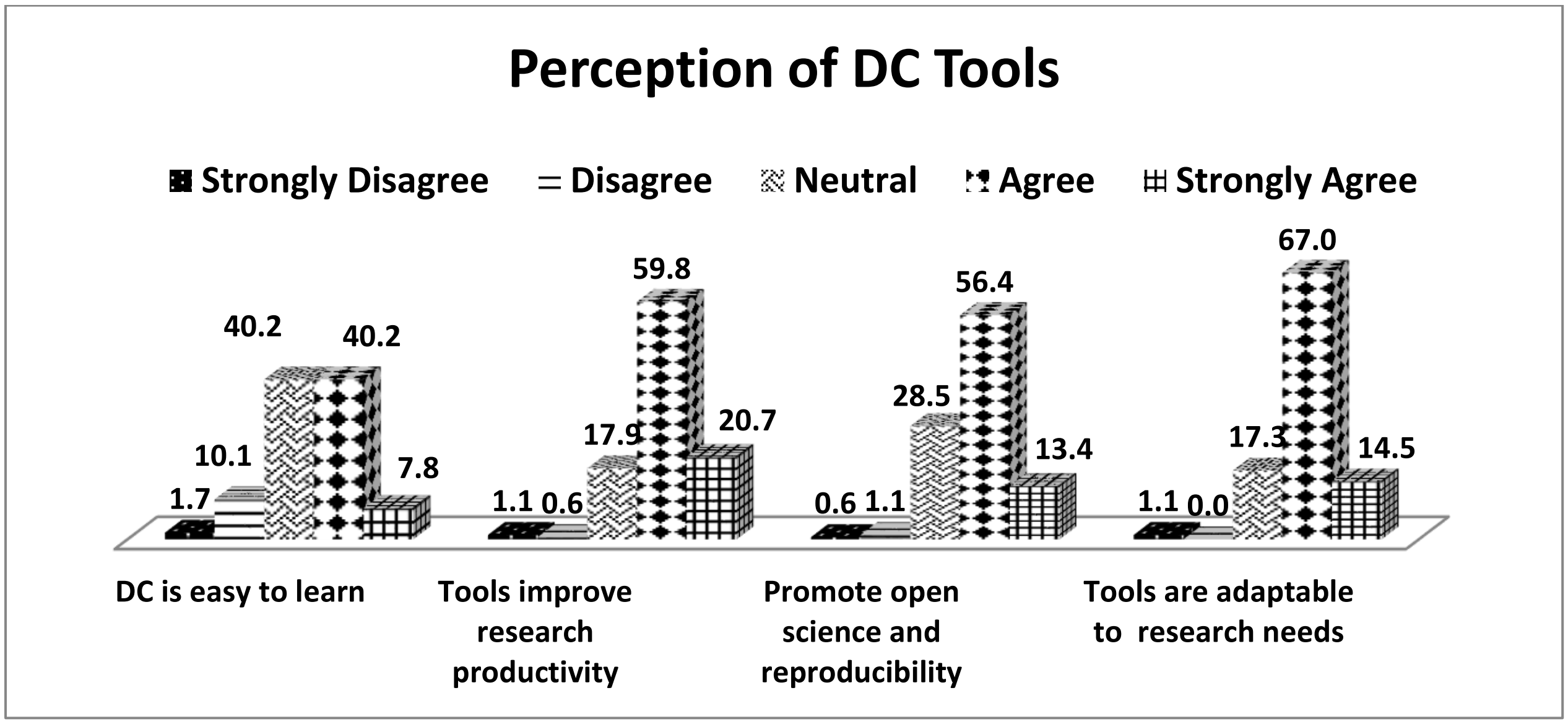

**Figure 1: Perception of DC Tools**

It is evident from the figure 1 that 40.2% researchers each have “Neutral” and “Agree” opinion in the statement of DC tools is easy to learn. At the same time 59.8% researchers ‘agree’ that DCT improve my research productivity. 56.4% researchers “agree” that DCT helps to promote open science and reproducibility. 67.0% researchers “agreed” that DCT are adaptable to their research needs.

Figure 1 makes clear that 40.2% of researchers both had "Neutral" and "Agree" opinions about the data statement that carpentry is an easy skill to master. Additionally, 59.8% of researchers "agree" that DCT increases their productivity as a researcher. 56.4% of researchers "agree" that DCT supports reproducibility and open science. Researchers "agreed" that DCT can be adjusted to meet their study goals in 67.0% of cases **(Objective 2).**

#### 6.2.4.1 Hypotheses

##### 6.2.4.1.1 Opinion regarding the perception of DC Tools of all statements are equal to average level

**Table 6: Perception of DC Tools (One-Sample T test)**

| Statements | Mean | SD | t value | p value |
|---|---|---|---|---|
| DC is easy to learn | 3.42 | 0.84 | 6.76 | 0.000** |
| Tools improve research productivity | 3.98 | 0.72 | 18.4 | 0.000** |
| Promote open science and reproducibility | 3.81 | 0.69 | 15.63 | 0.000** |
| Tools are adaptable to research needs | 3.94 | 0.65 | 19.43 | 0.000** |

** denotes significant level at 1% level.

Since p value is less than 0.01, the null hypothesis is rejected at 1% level with regard to all the statements of perception of DC tools. Hence concluded that the opinion regarding all the statements of perception of DC tools is not equal to average level. Based on mean score, the opinion regarding the statements on DC tools perception is above average level.

**6.2.4.1.2 There is no significant difference between Gender with regard to all the statements of DC Tools perception**

**Table 7: Gender v/s Perception of DC Tools (Independent-Samples T Test)**

| Statements | Male | | Female | | t value | p value |
|---|---|---|---|---|---|---|
| | Mean | SD | Mean | SD | | |
| DC is easy to learn | 3.77 | 0.81 | 3.25 | 0.81 | 4.03 | 0.000** |
| Tools improve research productivity | 4.13 | 0.59 | 3.91 | 0.76 | 2.01 | 0.046* |
| Promote open science and reproducibility | 3.88 | 0.74 | 3.77 | 0.67 | 1 | 0.317 |
| Tools are adaptable to research needs | 4.10 | 0.57 | 3.86 | 0.67 | 2.41 | 0.17 |

** denotes significant level at 1% level *denotes significant level at 5% level

Since p value is less than 0.01, the null hypothesis is rejected at 1 % level with regard to 'DC is easy to learn'. Hence there is significant difference between male and female R/S with regard to the statement 'DC is easy to learn'. Based on mean score, the male R/S (3.77) have better opinion on 'DC is easy to learn' than female (3.25). The result could point out that males possess a greater level of familiarity, prior experience, or confidence with technical tools, resulting in a heightened perception of ease. Women may encounter obstacles such as restricted exposure or diminished self-esteem in technical learning settings.

Since p value is less than 0.05, the null hypothesis is rejected at 5% level with regard to 'Tools improves research productivity'. Hence there is significant difference between male and female R/S with respect to the statement 'Tools improves research productivity'. Based on mean score, male have better opinion (4.13) on 'Tools improves research productivity' than female (3.91). Both genders recognize the productivity advantages of these tools; however, the marginally higher mean for males may be linked to a more effective integration of these tools into their research workflows or varying perceptions of their usefulness.

There is no significant difference between male and female R/S with respect to 'Promote open science and reproducibility' and 'Tools are adaptable to research needs'. Both males (3.88) and females (3.77) exhibit comparable perceptions concerning the role of DC tools in fostering open science and reproducibility. Moreover, the adaptability of these tools to research needs is rated by males at 4.10 and by females at 3.86. The p-value is greater than 0.05, indicating no statistically significant difference between the groups. Hence the null hypothesis is accepted at 5% level with regard to 'Promote open science and reproducibility' and 'Tools are adaptable to research needs'.

The 'Promote open science and reproducibility' result highlights a common comprehension among genders regarding the function of these tools in improving transparency and reproducibility in research, likely due to the universal nature of these elements, which are not affected by gender

Simultaneously the 'Tools are adaptable to research needs' results reveal that difference may arise from variations in how males and females approach or perceive technical solutions in research. Males may have more confidence in customizing or applying these tools to diverse research contexts

#### 6.2.4.1.3 Level of Perception of DC Tools are equally distributed among Research Scholars

**Table 8: Level of Perception of Data Carpentry Tools (Chi-Square goodness of fit)**

| Level of Perception | Frequency | Percentage | chi-square | P value |
|---|---|---|---|---|
| Low | 60 | 33.5 | 27.229 | 0.000** |
| Moderate | 31 | 17.3 | | |
| High | 88 | 49.2 | | |

Since p value is <0.01, the null hypothesis is rejected at 1% level. Hence the level of perception of DC tools is not equally distributed among R/S. Based on percentage, the majority of the research scholars belong to high level (49.2%).

The frequency and percentage show that out of 179 R/S, 49.2% belong to the "high" level of perception, while 33.5% are in the "low" category, and 17.3% are in the "moderate" category. Chi-Square value (27.229) represents the degree to which the observed distribution of perception levels differs from an expected equal distribution.

Since the p-value is less than 0.01, the null hypothesis that "the level of perception of DC tools is equally distributed among research scholars" is rejected at the 1% significance level.

The chi-square test goodness of fit compares low, moderate, and high frequency observations to the anticipated frequencies if perception levels were distributed equally. The significant finding shows that the observed frequency differences are too great for random chance. This illustrates that research experts' perceptions fluctuation. DC tools are seen differently by research researchers, with a large share having a "high" assessment. This shows that focused interventions like training or capacity-building programs may improve "low" and "moderate" perception.

Because research scholars take part with data-intensive activities, individuals having access to seminars, training programs, or self-learning opportunities have high perception. Researchers are required to be proficient in data collecting, cleaning, analysis, and visualization technologies, which increases awareness and competence. Scholars who recognize DC tools' significance for their study are more likely to see them favorably, resulting in the "high" perception level. Due to learning new technologies or a lack of perceived immediate value, those with reduced exposure or interest may fall into the "low" or "moderate" categories.

**6.2.4.1.4 There is no association between research scholars faculty of subjects and level of perception of DC Tools.**

**Table 9: Faculty of subject v/s Level of Perception of Data Carpentry Tools (Chi-Square Test)**

| Faculty of Subjects | Level of Perception of DC Tools | | | Total | Chi-square | p-value |
|---|---|---|---|---|---|---|
| | Low | Moderate | High | | | |
| **Science** | (29.2%) | (17.7%) | (53.1%) | (100%) | 1.85 | 0.397 |
| | [46.7%] | [54.8%] | [58%] | [53.6%] | | |
| **Social Science** | (38.6%) | (16.9%) | (44.6%) | (100%) | | |
| | [53.3%] | [45.2%] | [42%] | [46.4%] | | |
| **Total** | (33.5%) | (17.3%) | (49.2%) | (100%) | | |
| | [100%] | [100%] | [100%] | [100%] | | |

*The values within ( ) refers to Row Percentage.

** The values within [ ] refers to Column Percentage.

The Chi-square analysis examines the relationship between the faculty of subjects (Science vs. Social Science) and the level of perception of DC tools (low, moderate, high). A chi-square test is used to test for association.

Since p value is >0.05, the null hypothesis is accepted at 5% level of significance. Hence concluded that there is no association between faculty of subjects and level of perception of DC tools.

Based on Row percentage, perception of DC tools of Science faculty R/S have 29.2% at low level, 17.7% at moderate level and 53.1% at high level. Whereas, that of Social Science faculty R/S have 38.6% at low level, 16.9% at moderate level and 44.6% at high level.

The Chi-Square Value (1.85) and p-value (0.397) indicate that perception levels across faculties are likely related to random factors rather than systematic differences across Science and Social Science research experts.

Due to more exposure to data-intensive scientific research methods, science R/S are often more confident or proficient in utilizing these tools, according to faculty percentage study. Many Social Science researchers have a high perception, although many remain low than Science R/S. This may indicate less technical data use or less computational tool training in Social Science.

In relative distribution across faculties, more Science scholars had moderate and high perceptions, supporting the hypothesis that their academic background might enable them to apply these technologies.

The chi-square test indicates no correlation between the subject faculty and perception. The differences are not substantial enough to indicate that faculty type affects perception systematically. Science scholars had higher views (53.1% high level) than Social Science scholars (44.6%). Science fields commonly use quantitative data analysis, programming, and computational tools, which fit DC tools. Research training for Science R/S may include more seminars or tools. Social Science R/S may see these techniques as unrelated to their qualitative or mixed-methods research.

Focused Social Science training programs should emphasize DC tools use in qualitative and mixed-methods research to fill the disparity. Workshops for Social Science academics help close the perception barrier and improve acceptance.

### 6.2.5 Level of Usefulness of Data Carpentry Tools

DC tools are meant to provide researchers with the skills and methodologies they need to efficiently manage, analyze, and share data.

**Table 10: Usefulness of DC tools**

| Opinion | No. of Researchers | % |
|---|---|---|
| Not Useful | 3 | 1.7 |
| Slightly Useful | 4 | 2.2 |
| Neutral | 26 | 14.5 |
| Useful | 105 | 58.7 |
| Very Useful | 41 | 22.9 |
| **Total** | **179** | **100.0** |

Table 10 shows that more than half of respondents (58.7%) found DC tools are useful and with 22.9% considering them very useful. 105 respondents (14.5%) are neutral, with 2.2% finding DC tools slightly useful and 1.7% finding them not useful.

### 6.2.6 Challenges in using Data Carpentry Tools

In majority of the studies the respondents claimed various types of obstacles while utilizing tools.

**Table 11: Challenges in using Data Carpentry Tools**

| Challenges | No. of Researchers | % |
|---|---|---|
| Lack of Technical Skills | 44 | 24.6 |
| Limited Access to Training | 72 | 40.2 |
| Time Constraints | 31 | 17.3 |
| Lack of Institutional Support | 31 | 17.3 |

*Multiple Choice

The leading cause of challenging, cited by almost half (40.2%) of respondents, was a lack of training opportunities, followed by a lack of technical skills (24.6%). However, 17.3% identified time restrictions and a lack of institutional support as their primary problems. **(Objective 4)**

### 6.2.7 Challenges in using Specific Data Carpentry Tools

The technologies are transforming every second. The majority of the tools we use have changed, and new user-friendly tools are being introduced daily. However, students will need more time to get training on that software, which makes it challenging to utilize the tools effectively.

**Table 12: Challenges in using Specific Data Carpentry Tools**

| Data Carpentry Specific Tools | Usage | % | Challenges | % |
|---|---|---|---|---|
| R Programming | 89 | | 89 | 60.3 |
| Python Programming | 58 | | 38 | 26.3 |
| SQL | 45 | | 5 | 4.5 |
| Git (Version Control) | 14 | | 16 | 8.9 |
| | **Total** | | **179** | **100.0** |

Table 12 shows that more than half of respondents (60.3%) consider R Programming is the most challenging tool, followed by Python Programming (26.3%). 8.9% of respondents trust Git (Version Control) is the most difficult technology, while 4.5% find SQL

### 6.2.8 Additional Support For Better Use of Data Carpentry Tools

Several types of additional support have been claimed by the researchers regarding the improved utilization of DC tools.

**Table 13: Additional Support for Better Use of Data Carpentry Tools**

| Additional Support | No. of Researchers | % |
|---|---|---|
| More Workshops and Training | 77 | 43.0 |
| Discipline-Specific Tutorials | 37 | 20.7 |
| Access to Mentors | 46 | 25.7 |
| Online Resources | 53 | 29.6 |

*Multiple Choice

Researchers, particularly beginners, may get organized help via workshops and training (43%). Users who prefer self-paced learning or lack in-person instruction value online tutorials, documentation, and video guides (29.6%). Users may solve obstacles faster with personalized instruction, troubleshooting, and practical advice from mentors (25.7%). DC solutions may be tailored to users' requirements using discipline-specific resources (20.7%).

## 7. Findings

1. DC has a major role in increasing the productivity of a researcher.

2. Male research scholars have a higher percentage in awareness of tools as well as integrating them in their research.
3. Research scholars have a higher perception in the area of data carpentry.
4. Scholars with science background have more perception about DC tools than scholars with social science background.
5. Lack of training programs/ workshops are the major challenge faced by the researcher.
6. Researchers are enthusiastic to share knowledge on DC tools to their peers.

## 8. Conclusion

The study found that researchers have a positive attitude towards DC tools and incorporating them into research, male scholars are more aware, research scholars have a higher perception of data mining but it is not evenly distributed, science back ground researchers have more perception than social science students, and scholars are more positive about sharing knowledge. Four hypotheses were generated at the beginning of the study; some of these assumptions were proven and rejected using statistical software. DC tools are valuable to most researchers, but lack of training camps and workshops is limiting their adoption. The research community suggests institution-wide seminars and university-level initiatives to empower researchers with DC skills.